\documentclass[runningheads]{llncs}
\usepackage{graphicx}
\usepackage{textcomp}
\usepackage{comment}
\usepackage{todonotes}
\usepackage{color}
\usepackage{array}
\newcolumntype{x}[1]{>{\centering\arraybackslash\hspace{0pt}}p{#1}}

\usepackage{multirow}

\begin{document}
\title{Improved AI-based Segmentation of Apical and Basal Slices from Clinical Cine CMR}  

\author{
Jorge Mariscal-Harana \inst{1} \and
Naomi Kifle \inst{1} \and
Reza Razavi \inst{1,2} \and
Andrew P. King \inst{1} \and
Bram Ruijsink \inst{1,2,3,*} \and
Esther Puyol-Ant\'on \inst{1,*}
}
\authorrunning{J Mariscal-Harana et al.}   
\titlerunning{Improved AI-based Segmentation of Apical and Basal CMR Slices}
\institute{
School of Biomedical Engineering \& Imaging Sciences, King's College London, UK. \and
Department of Adult and Paediatric Cardiology, Guy's and St Thomas' NHS Foundation Trust, London, UK. \and
Department of Cardiology, Heart and Lung Division, University Medical Center Utrecht, Utrecht, The Netherlands.
}
\maketitle              
*Shared last authors

\begin{abstract} 
Current artificial intelligence (AI) algorithms for short-axis cardiac magnetic resonance (CMR) segmentation achieve human performance for slices situated in the middle of the heart. However, an often-overlooked fact is that segmentation of the basal and apical slices is more difficult. During manual analysis, differences in the basal segmentations have been reported as one of the major sources of disagreement in human interobserver variability. In this work, we aim to investigate the performance of AI algorithms in segmenting basal and apical slices and design strategies to improve their segmentation. We trained all our models on a large dataset of clinical CMR studies obtained from two NHS hospitals (n=4,228) and evaluated them against two external datasets: ACDC (n=100) and M\&Ms (n=321). Using manual segmentations as a reference, CMR slices were assigned to one of four regions: non-cardiac, base, middle, and apex. Using the ‘nnU-Net’ framework as a baseline, we investigated two different approaches to reduce the segmentation performance gap between cardiac regions: (1) non-uniform batch sampling, which allows us to choose how often images from different regions are seen during training; and (2) a cardiac-region classification model followed by three (i.e. base, middle, and apex) region-specific segmentation models. We show that the classification and segmentation approach was best at reducing the performance gap across all datasets. We also show that improvements in the classification performance can subsequently lead to a significantly better performance in the segmentation task.
\end{abstract}

\section{Introduction}
Segmenting the ventricles and the myocardium from cardiac magnetic resonance (CMR) images allows cardiologists to assess cardiac function, thus guiding diagnosis, prognosis and treatment of cardiac disease \cite{Peng2016}. However, for both ventricles, the selection and segmentation of the basal slice are major determinants of variability in the assessment of cardiac function \cite{Caudron2012}. For most of the slices, the ventricular blood pool is a circular shape whose border is well defined by the presence of a ring of myocardial muscle. However, in the basal slices, the myocardial muscle does not completely enclose the blood pool and its shape can be variable and non-circular due to the presence of the outflow tract and the atrio-ventricular and ventriculo-arterial valves. Moreover, through-plane motion of the cardiac structures over the cardiac cycle further challenges reproducible segmentations. In the apex, the myocardial rim is angulated with respect to the short-axis imaging plane, which means that the borders are less well defined. Incorrect segmentation of these two areas of the ventricle contributes to errors in measurement of cardiac function. Due to the longitudinal motion of the heart during contraction, the basal slices contribute greatly to estimates of stroke volume (i.e. the ejected volume of blood during ventricular contraction).

Deep learning (DL) models have achieved human-level performance for automatic segmentation of short-axis cine CMR images \cite{Bernard2018}, and have been recently proposed as a means for the automated characterisation of cardiac function \cite{Ruijsink2020}. However, model performance is not uniform throughout the heart, with basal and apical regions posing a greater challenge than middle regions \cite{Bernard2018}. This is likely to be caused by the inherent difficulty in segmenting basal and apical regions, as well as the higher variability of manual segmentations.

In this paper, to the best of our knowledge, we perform the first statistical analysis of the performance of DL-based segmentation across the different sections of the heart. We discover significant differences between middle slices and both basal and apical slices after training a state-of-the-art segmentation model on a large clinical dataset (n=4,228) and testing it on three independent test datasets. We propose two approaches to bridge the performance gap between cardiac regions, resulting in significantly improved segmentations for basal and apical slices with a reduced error spread.

\section{Methods}
\label{sec:methods}
\begin{figure}[ht]%
    \centering
    \includegraphics[trim={0 2cm 0 2cm}, clip, width=\textwidth]{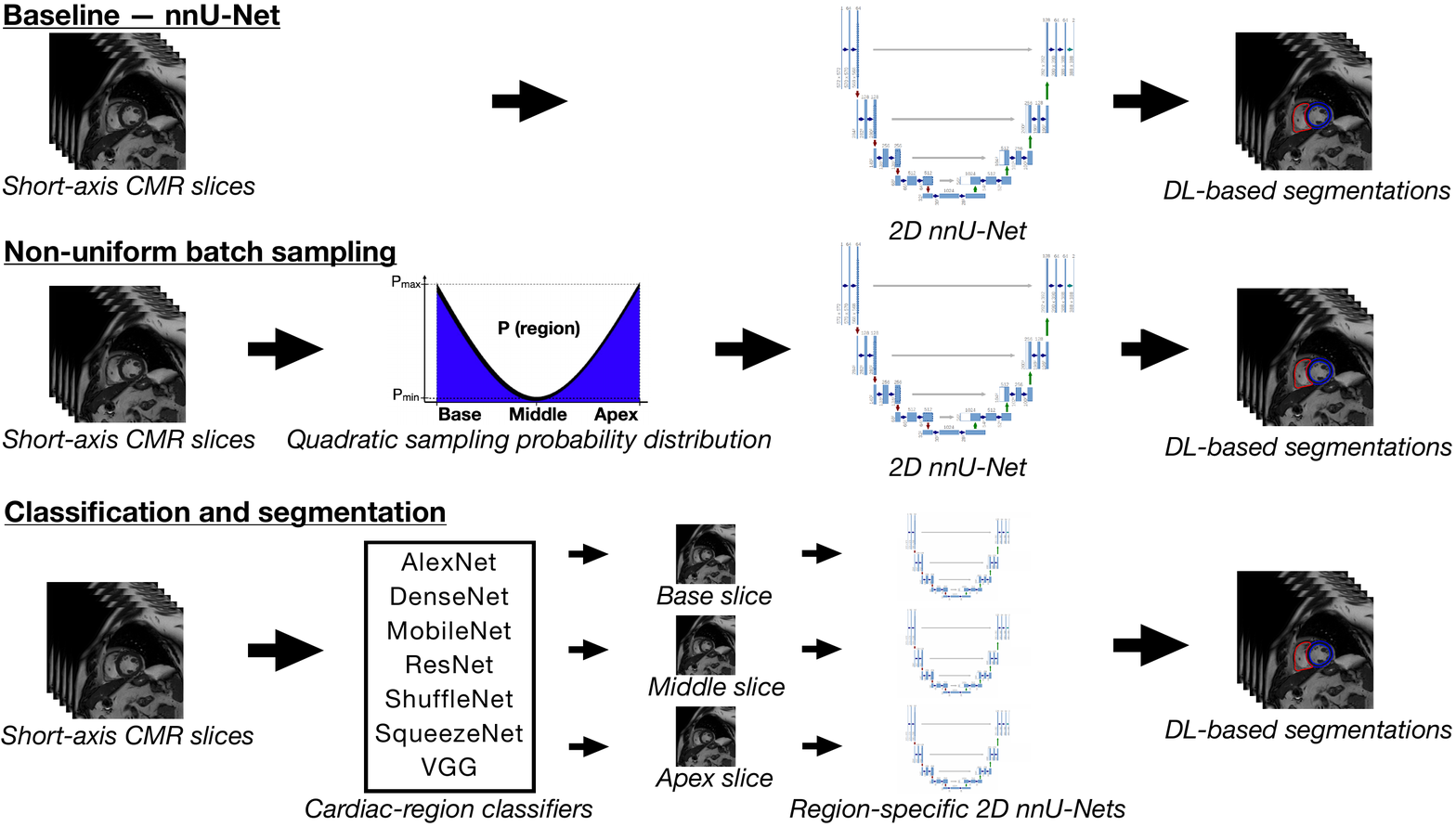}
    \caption{Baseline and proposed approaches to reduce the performance gap between cardiac regions. The baseline corresponds to the original nnU-Net framework trained on the NHS dataset. (1) Non-uniform batch sampling approach using a quadratic probability distribution. (2) Cardiac-region classifier followed by region-specific segmentation models.}
    \label{fig:overview}%
\end{figure}

To investigate whether DL-based automated CMR segmentation suffers from significant performance differences between cardiac regions, we trained a baseline segmentation model using a large dataset of clinical CMR images (n=4,228). We subsequently investigated two potential approaches to reduce the performance gap based on (1) non-uniform batch sampling and (2) a combination of a cardiac-region classification model and three region-specific segmentation models. The baseline and the proposed approaches are illustrated in Fig. \ref{fig:overview}.

\noindent
\textbf{Segmentation model}: 
We used the ‘nnU-Net’ framework (2D U-Net) \cite{Isensee2020} to segment the left and right ventricular blood pools (LVBP and RVBP, respectively) and the left ventricular myocardium (LVM) from end-diastole (ED) and end-systole (ES) short-axis CMR stacks. Details of the training and test datasets and experimental setup are provided in Section \ref{sec:results}.\\
\textbf{Baseline:} We trained the nnU-Net model on the NHS clinical CMR dataset. This approach is considered the baseline since its performance is comparable to that of other state-of-the-art CMR segmentation methods \cite{Bernard2018}.\\
\textbf{Approach (1) - Non-uniform batch sampling:} We modified nnU-Net’s data loader to allow for a “region-aware” non-uniform batch sampling. This means that, for each mini-batch during training, slices from the apex and the base are more likely to be selected than middle slices. This is achieved via a quadratic probability distribution whose lower and upper limits determine the minimum and maximum sampling probabilities, respectively. This approach does not affect the probability of non-cardiac slices being selected.\\
\textbf{Approach (2) - Classification and segmentation:} We fine-tuned seven state-of-the-art DL-based classification models to determine which cardiac region each slice belonged to: AlexNet \cite{NIPS2012_c399862d}, DenseNet121 \cite{Huang2017}, MobileNetV2 \cite{Sandler2018}, ResNet18 \cite{He2016},  ShuffleNet V2 \cite{Ma2018}, SqueezeNet \cite{Iandola2016}, and VGG Net-A \cite{Simonyan15} (with batch normalisation). Since there was a class imbalance in terms of cardiac regions -- with non-cardiac and middle slices being more common than basal and apical ones -- we investigated the use of a weighted loss function for classification. However, this resulted in a decrease in performance across all models, thus we used a non-weighted loss function instead. We also trained three region-specific segmentation models using slices from the corresponding cardiac regions only. To ensure that the same number of images would be seen by each model, we used nnU-Net’s default hyperparameters. Finally, we combined the best-performing classifier with the individual segmentation models for our predictions. Note that slices classified as non-cardiac were analysed using the baseline model.\\
\textbf{Evaluation metrics:} Classification performance was evaluated using precision and recall for each slice. To evaluate segmentation performance, we used the Dice similarity coefficient (DSC) between manual segmentations and model predictions for the LVBP, RVBP, and LVM. Since the number of segmented slices was variable, DSC distributions from basal to apical slices were calculated via linear interpolation (n=101) to evaluate regional segmentation performance. 
Finally, we assessed improvements in segmentation performance between the baseline and the proposed approaches via the differences in the mean and the standard deviation (SD) of the DSC.

\section{Materials and Experiments}
\label{sec:results} 
\textbf{Datasets:} The baseline and proposed approaches were trained and evaluated using manual segmentations from a dataset of clinical CMR scans (n=4,684) acquired at two NHS hospitals and including the full spectrum of cardiac disease phenotypes. All approaches were evaluated on a randomly selected subset from the NHS (n=456) dataset, and on two external datasets of clinical CMRs: ACDC \cite{Bernard2018} (n=100) and M\&Ms \cite{Campello2021} (n=321). All datasets consisted of ED and ES short-axis cine CMR images. For the NHS dataset, left-ventricular endocardial and epicardial borders and RV endocardial borders were manually segmented using the commercially available cvi42 CMR analysis software (Circle Cardiovascular Imaging Inc., Calgary, Alberta, Canada, version 5.10.1). Although working with data acquired at different spatial resolutions and MR parameters is challenging, nnU-Net's resampling during preprocessing helps to mitigate its impact on the results.\\
\textbf{Training:} For all approaches, we split the NHS dataset into random training, validation, and test splits with 3382, 846, and 456 cases, respectively. We found that the single fold performance of the baseline segmentation model was similar to that of the five-fold cross-validation, so all models used in this work were trained on a single random fold. All segmentation models were trained for 1,000 epochs with 250 random mini-batches of size 32 per epoch, and optimised using nnU-Net’s default configuration: stochastic gradient descent (initial learning rate of 0.01 with a polynomial learning rate policy, Nesterov momentum of 0.99, and weight decay of 3e-5). Classification models were trained for 1,000 epochs with the Adam optimiser (initial learning rate of 5e-4). All models were trained on a NVIDIA TITAN RTX GPU. The approximate training times for each segmentation and classification models were 24 and 42 hours, respectively.

We compared the nnU-Net baseline to the two proposed approaches using metrics of segmentation performance. For each short-axis CMR stack, slices containing segmentations were considered to belong to the base/middle/apex cardiac regions following a 20\%/60\%/20\% split, respectively. The ratio of stratification was based on cardiac anatomical knowledge and clinical experience.

\noindent
\textbf{Experiment 1 - Performance gap assessment:} We first looked at the regional differences in segmentation performance for the baseline model. Table \ref{table:experiment1} shows the mean and SD of the DSC for each cardiac region and segmentation label. Statistically significant differences between the base-middle and apex-middle regions were found using a Welch unequal variances \textit{t}-test for all labels and test datasets, except for the ACDC dataset, which did not show significant differences in basal LVBP and RVBP.

\begin{table}[ht]
\caption{Mean (and SD) of the DSC (\%) for each cardiac region and segmentation label. Asterisks indicate statistically significant differences in the mean DSC between the base or the apex and the middle region (p$<$0.01).}
\centering
\begin{tabular}{p{1.5cm} x{1.5cm}x{2.5cm}x{2.5cm}x{2.5cm}}
\multirow{2}{*}{Dataset} & \multirow{2}{*}{Label} & \multicolumn{3}{c}{Cardiac regions}             \\
\cline{3-5}
                         &                        & Base           & Middle        & Apex           \\
\hline
\multirow{3}{*}{NHS}     & LVBP                   & 87.53* (23.29) & 94.93 (5.62)  & 81.08* (24.86) \\
                         & LVM                    & 79.82* (19.34) & 87.22 (6.04)  & 70.18* (24.96) \\
                         & RVBP                   & 76.25* (28.90) & 90.72 (9.57)  & 73.24* (27.03) \\
\hline
\multirow{3}{*}{M\&Ms}   & LVBP                   & 84.41* (20.38) & 93.63 (6.73)  & 89.86* (16.47) \\
                         & LVM                    & 70.23* (29.13) & 86.46 (8.08)  & 75.25* (25.84) \\
                         & RVBP                   & 76.60* (26.56) & 90.11 (11.84) & 81.62* (25.36) \\
\hline
\multirow{3}{*}{ACDC}    & LVBP                   & 94.42 (7.76)   & 94.28 (6.87)  & 82.86* (23.47) \\
                         & LVM                    & 84.01* (17.59) & 87.66 (9.34)  & 64.99* (33.28) \\
                         & RVBP                   & 90.23 (16.02)  & 88.54 (16.22) & 67.42* (33.97)\\
\hline
\end{tabular}
\label{table:experiment1}
\end{table}

\begin{figure}[ht]%
    \centering
    \includegraphics[width=1.0\textwidth,trim={2.3cm 0 2.3cm 0},clip]{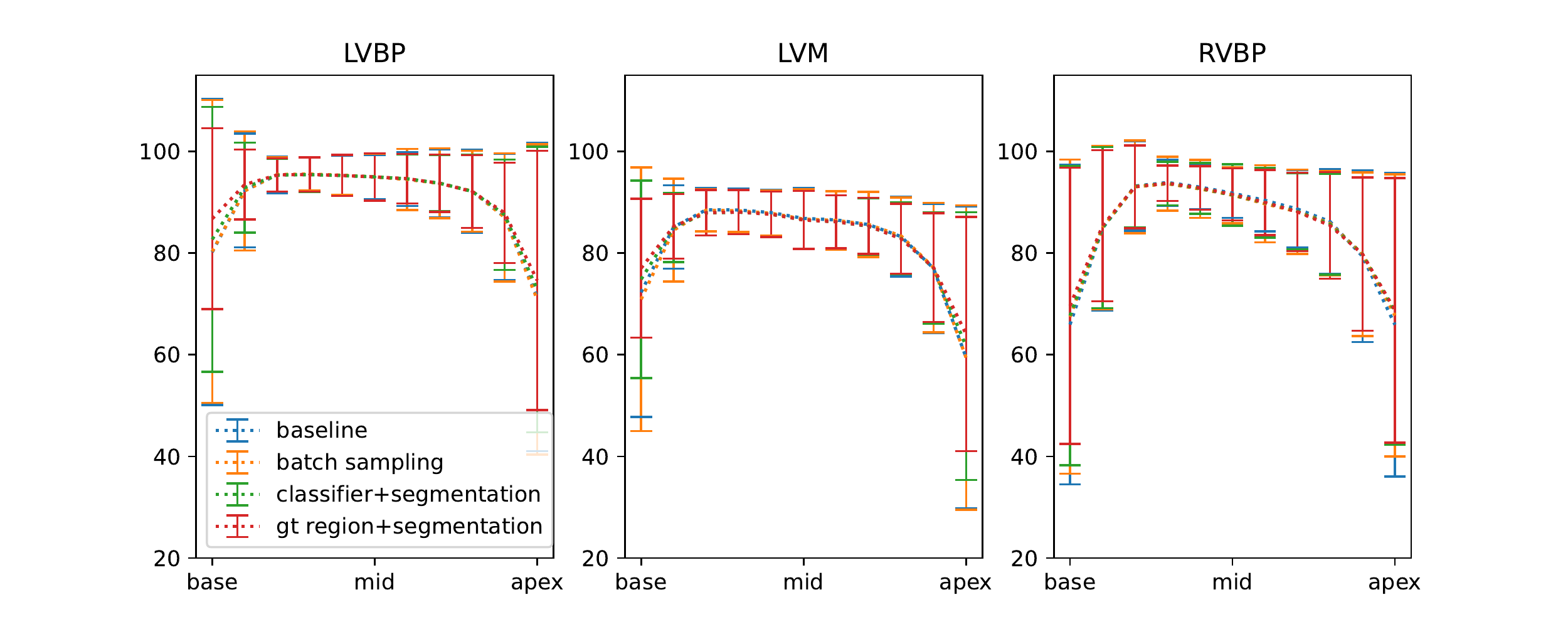}
    \caption{Interpolated DSC distribution across the heart for the LVBP (left), LVM (middle), and RVBP (right) for the baseline and proposed methods for the NHS test cases. Blue: baseline, orange: approach (1); green: approach (2); red: upper-bound performance of approach (2).}
    \label{fig:best_NHS_4}
\end{figure}

\noindent
\textbf{Experiment 2 - Reducing the performance gap:} Given the results from the first experiment, we focused on improving the segmentation performance for basal and apical slices with the two proposed approaches: (1) non-uniform batch sampling and (2) classification and segmentation. For approach (1), we tried two batch sampling levels - “low” and “high” - by setting the ratio of maximum to minimum sampling probabilities to 4 and 20, respectively. A higher level of batch sampling produced better results. For approach (2), we compared the seven classifiers according to the weighted-average precision and recall values for all cardiac regions in the NHS test cases (Table \ref{table:classifiers}). The best-performing classifier, DenseNet121, was combined with the region-specific segmentation models. Additionally, the ground truth cardiac regions were combined with the region-specific segmentation models to determine the upper-bound performance (i.e. equivalent to a perfect classifier) of this approach. Fig. \ref{fig:best_NHS_4} shows the interpolated DSC distributions for the baseline, the best-performing batch sampling approach, the classification and segmentation approach, and the ground truth region and segmentation. For the base and the apex, approach (2) produces the greatest performance improvements.

\begin{table}[ht] 
\caption{Mean precision and recall values for seven state-of-the-art classification models trained and tested on the NHS dataset.}  
\centering
\begin{tabular}{p{1.9cm} ccccccc}
                   & AlexNet & DenseNet & MobileNet & ResNet & ShuffleNet & SqueezeNet & VGG  \\
\hline
Precision {[}\%{]} & 87.54   & 91.72    & 89.56     & 90.94  & 90.33      & 87.62      & 91.02 \\
Recall {[}\%{]}    & 87.56   & 91.50    & 89.50     & 90.94  & 90.25      & 87.56      & 91.06 \\
\hline
\end{tabular}
\label{table:classifiers}
\end{table}

Finally, Table \ref{table:DSC_differences} shows the differences in the mean and the SD of the DSC between the baseline and the proposed approaches for all datasets. Paired Student's \textit{t}-tests were performed to determine statistically significant differences between the means of the baseline and the proposed approaches. The results show that the classification and segmentation approach outperforms both the baseline and the non-uniform batch sampling approach across all segmentation labels and datasets. We observed that segmentation performance for the background label (i.e. not LVBP, LVM, or RVBP) was uniform across all slices and similar for all proposed methods.

\begin{table}[ht] 
\caption{Differences in DSC mean (and SD) values between the baseline and the proposed approaches for the base and the apex. Positive values indicate improvements over the baseline (i.e. a higher mean and a lower SD). The largest improvement for each label and cardiac region is highlighted in bold. Asterisks indicate statistically significant differences between the means (p$<$0.01).}
\centering
\begin{tabular}{l x{2.5cm}x{1.2cm}x{1.2cm}x{1.2cm}x{1.2cm}x{1.2cm}x{1.2cm}}
\multirow{2}{*}{Dataset} & \multirow{2}{*}{Approach}     & \multicolumn{2}{c}{LVBP}                       & \multicolumn{2}{c}{LVM}                       & \multicolumn{2}{c}{RVBP}                       \\
\cline{3-8}
                         &                               & Base                   & Apex                  & Base                  & Apex                  & Base                   & Apex                  \\
\hline
\multirow{2}{*}{NHS}     & Batch sampling                & -0.06 (0.01)           & -0.27 (-0.29)         & -0.91* (-1.53)        & 0.09 (-0.22)          & 0.89* (0.67)           & 1.03* (1.82)          \\
                         & Classification + segmentation & \textbf{1.33* (3.35)}  & \textbf{0.78* (1.91)} & \textbf{0.88* (3.90)} & \textbf{0.97 (2.89)}  & \textbf{1.04* (1.74)}  & \textbf{1.58* (3.21)} \\
\hline
\multirow{2}{*}{M\&Ms}   & Batch sampling                & 0.42 (1.39)            & -0.21 (-0.13)         & 0.28 (0.34)           & -1.33* (-1.37)        & 1.61* (3.30)           & \textbf{0.76 (2.38)}  \\
                         & Classification + segmentation & \textbf{0.85* (2.59)}  & \textbf{0.74* (2.83)} & \textbf{5.71* (8.02)} & \textbf{5.53* (9.38)} & \textbf{2.44* (4.57)}  & 0.02 (2.58)           \\
\hline
\multirow{2}{*}{ACDC}    & Batch sampling                & -0.28 (-0.00) & -0.75 (-0.46)         & -0.88 (-0.94)         & 0.46 (0.92)           & -0.61 (-0.57) & 3.03* (3.78)          \\
                         & Classification + segmentation & -0.46 (-0.08)          & \textbf{0.51 (3.36)}  & \textbf{1.88* (9.34)} & \textbf{7.60* (9.19)} & -1.37* (-0.78)         & \textbf{5.74* (6.87)} \\
\hline
\end{tabular}
\label{table:DSC_differences}
\end{table}

\section{Discussion}
\label{sec:discussion}
To the best of our knowledge, this paper has presented the first statistical analysis of the performance of DL models for cardiac segmentation across different regions of the heart. We have shown that there are significant segmentation performance differences between the middle of the heart and both the apex and the base in three test datasets. In the ACDC dataset, however, the CMR volumes were truncated and did not include the full basal portion of the heart, which explains the lack of significant differences between the base and the middle for the LVBP and RVBP. These findings are in line with experience from clinicians in daily practice. Although human-level segmentation performance has been achieved by DL-based models overall, these regional differences are critical for clinical evaluation since errors in segmentation at the base and apex can impact subsequent estimates of functional metrics such as stroke volume and ejection fraction.

We have proposed two approaches to increase the DL-based segmentation performance for the base and the apex and compared them to a nnU-Net baseline. The first approach (non-uniform batch sampling) allowed us to modify how often CMR slices from different cardiac regions were seen by the model during training. The second approach (classification and segmentation) combined a cardiac-region classifier with region-specific segmentation models with the aim of having specialised models for each cardiac region. Our results show that both approaches improve performance in the basal and apical regions, although only the classification and segmentation approach produced significantly better results across all labels and datasets. This shows that the characteristics of basal and apical regions seem to differ enough from the middle slices to regard them as different tasks with respect to the segmentation model. Furthermore, the results for the upper-bound performance of apprach (2) are promising, as they indicate that further segmentation improvements could be achieved by improving the performance of the cardiac-region classifiers.

In conclusion, in this work we have highlighted the uneven segmentation performance of DL-based models across the heart and proposed two approaches to bridge this performance gap. We have shown that combining a classifier with region-specific segmentation models significantly improves segmentation performance at the base and the apex. In the future, motivated by the current results, we aim to train deeper classification models to improve our current cardiac-region classification performance and to investigate a simultaneous classification and segmentation approach. Additionally, we will analyse how an improved segmentation performance affects the estimation of cardiac function metrics such as diastolic and systolic ventricular volumes, ejection fraction, and left ventricular mass.

\subsubsection*{Acknowledgements}
This work was supported by the EPSRC (EP/P001009/1 and the Advancing Impact Award scheme of the Impact Acceleration Account at King's College London) and the Wellcome EPSRC Centre for Medical Engineering at the School of Biomedical Engineering and Imaging Sciences, King's College London (WT 203148/Z/16/Z).

\bibliographystyle{splncs04}
\bibliography{MAIN}

\end{document}